\documentclass[prl,amssymb,twocolumn,tightenlines,showpacs]{revtex4-1}

\usepackage{graphicx}
\usepackage{amsmath}
\usepackage{verbatim}
\usepackage{subfigure}
\usepackage{epstopdf}
\usepackage{color}
\usepackage{natbib}

\definecolor{mygrey}{gray}{0.35}
\definecolor{myblue}{rgb}{0.2,0.2,0.8}
\definecolor{myzard}{cmyk}{0,0,0.05,0}
\definecolor{mywhite}{rgb}{1,1,1}
\definecolor{myred}{rgb}{1,0.,0.3}

\newcommand{\be}{\begin{equation}}
\newcommand{\ee}{\end{equation}}

\usepackage[colorlinks=true,citecolor=myblue,linkcolor=myred,urlcolor=myblue]{hyperref}

\begin{document}

\title{Quantum tomography of photon states encoded in polarization and picosecond time-bins
}
\author{Y. Pilnyak}
\affiliation{Racah Institute of Physics, Hebrew University of Jerusalem,
Jerusalem 91904, Israel}
\author{P. Zilber}
\affiliation{Racah Institute of Physics, Hebrew University of Jerusalem,
Jerusalem 91904, Israel}
\author{L. Cohen}
\affiliation{Racah Institute of Physics, Hebrew University of Jerusalem,
Jerusalem 91904, Israel}
\author{H. S. Eisenberg}
\affiliation{Racah Institute of Physics, Hebrew University of Jerusalem,
Jerusalem 91904, Israel}


\begin{abstract}
A single photon has many physical degrees of freedom (DOF) that can carry the state of a high-dimensional quantum system. Nevertheless, only a single DOF is usually used in any specific demonstration. Furthermore, when more DOF are being used, they are analyzed and measured one at a time. We introduce a two-qubit information system, realized by two degrees of freedom of a single photon: polarization and time. The photon arrival time is divided into two time-bins representing a qubit, while its polarization state represents a second qubit. The time difference between the two time-bins is created without an interferometer at the picosecond scale, which is much smaller than the detector's response time. The two physically different DOF are analyzed simultaneously by photon bunching between the analyzed photon and an ancilla photon. Full two-qubit states encoded in single photons were reconstructed using quantum state tomography, both when the two DOF were entangled and when they were not, with fidelities higher than 96\%.

\end{abstract}

\maketitle

\section{I. INTRODUCTION}
Entanglement is a non-trivial property that is part of quantum theory
\cite{EPR35}. It describes how seemingly separated quantum objects can only
be described as a single inseparable physical entity. In the most basic
example, two entangled particles can be used to demonstrate the non-locality
of quantum mechanics \cite{Bell64}. The first physical realization of
entanglement was witnessed using photons. This has been done by entangling the
photons' different degrees of freedom (DOF) such as position, linear and
angular momentum, polarization and arrival time \cite{Aspect81}. Entanglement
is not restricted to a pair of particles, but can be extended to more as
well \cite{GHZ90,Bouwmeester99}.

The efforts to achieve entangled states with higher number of particles originate from the
interest in quantum states of high dimensionality. It has been suggested that
adding DOF to a state to increase its dimensionality can be achieved not only
by adding particles, but also by using more intrinsic DOF of the same particles
\cite{Kwiat97,Barreiro05}. Such states of hybrid DOF have larger
dimensionality with fewer particles. Apart from the added dimensions, they
enable protocols that are otherwise hard or impossible to realize. An example for
such a protocol is a full Bell state measurement with linear optical elements
of polarized photons \cite{Kwiat98,Walborn03}, which is otherwise impossible
\cite{Vaidman99,Lutkenhaus99}.

As one of the most successful realizations of qubits, photons are widely used in experimental demonstrations of quantum information. Manipulation of photons is relatively easy, as different operations could be performed using linear optical elements. While usually the photon polarization degree of freedom is used to encode the quantum information, other DOF are available as possible realizations of qubits as well. Generation of time-frequency entanglement was demonstrated \cite{Brendel99,Simon05,Zavatta06}, with its advantages accounted also for quantum key distribution \cite{Tittel00}. Various recent demonstrations presented hybrid DOF with dimensionality of up to 10 qubits $(d=2^{10})$ \cite{Vallone07,Gao10,Lee12,Malik16}.

Combining the polarization and time-bin DOF of a photon has been suggested in a hybrid multi-photon scheme for a single-mode quantum computer, where information is stored in the temporal DOF, while the polarization DOF serves as a bus to route the qubits to their quantum logical gates \cite{Humphreys13}. One demonstration that encoded information only in short temporal differences has used the nonlinear process of sum-frequency generation with an intense shaped laser pulse to characterize the temporal information \cite{Donohue13}. Due to detectors that cannot discriminate short temporal differences, both of these demonstrations have used long delay lines in order to observe or implement the different time-bins and their desired operations.

Controlled coupling between the polarization DOF and the temporal DOF has also been used for another goal. By using birefringent crystals, the polarization and temporal DOF of a single photon have been entangled. As the small time differences were undetectable by the photon detectors, the temporal DOF was practically traced out, leading to the controllable depolarization process of the quantum information carried by the polarization DOF of a single photon \cite{Shaham11}.

In this work we present and demonstrate a scheme which encodes a two-qubit state in a single photon, using two of its DOF - polarization and time. While previous works have shown different methods to incorporate the time DOF, our scheme is simple, and does not require interferometrically stable delay loops. In addition, we present a measurement method which analyzes all of the DOF of a single photon simultaneously, including temporal differences that are too small to detect by current detectors.

\section{II. ENCODING PROCESS}
In order to encode two qubits in a single photon, the first qubit is encoded in the photon's polarization DOF. The horizontal $|h\rangle$ and the vertical $|v\rangle$ polarization states encode the logical $|0\rangle_p$ and $|1\rangle_p$ states. The second qubit is encoded in the arrival time-bin information. This is in principle a continuous DOF, that can be used as a finite discrete DOF with a predefined set of time-bin slots. We encode the logical $|0\rangle_t$ in the non-delayed $t=0$ time-bin $|0\rangle$ and the logical $|1\rangle_t$ in the $t=\tau$ delayed time-bin $|\tau\rangle$. For later use we define the linearly polarized state at $45^\circ$ to the horizon as the diagonal (anti-diagonal) $|p\rangle=\frac{1}{\sqrt{2}} \left( |h\rangle+|v\rangle \right)$ $\left(|m\rangle=\frac{1}{\sqrt{2}} \left( |h\rangle-|v\rangle \right) \right)$, and the right (left) circularly polarized states as $|r\rangle=\frac{1}{\sqrt{2}} \left( i|h\rangle+|v\rangle \right)$ $\left(|l\rangle=\frac{1}{\sqrt{2}} \left( |h\rangle+i|v\rangle \right) \right)$. As with the polarization, any superposition of the two time-bins is possible. Accordingly, we designate the following temporal superpositions: $|+\rangle=\frac{1}{\sqrt{2}} \left( |0\rangle+|\tau\rangle \right)$, $|-\rangle=\frac{1}{\sqrt{2}} \left( |0\rangle-|\tau\rangle \right)$, $|\times\rangle=\frac{1}{\sqrt{2}} \left( i|0\rangle+|\tau\rangle \right)$, and $|\div\rangle=\frac{1}{\sqrt{2}} \left( |0\rangle+i|\tau\rangle \right)$.

\begin{figure}[t]
\includegraphics[width=\columnwidth]{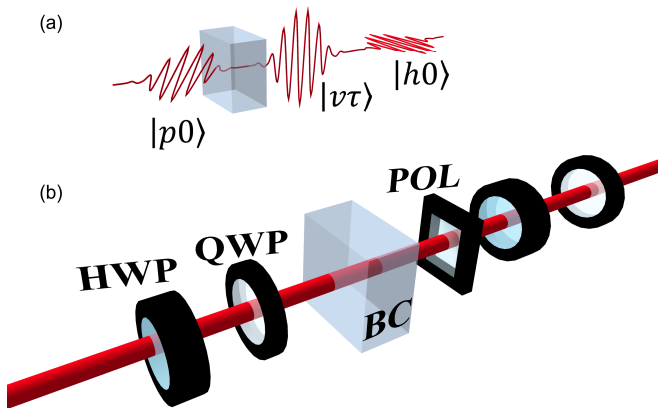}
\caption{\label{Fig1} Encoding polarization-time qubits on single photons. (a) Effect of birefringent crystal. (b) The photons travel from left to right, where their initial state is controlled by polarization rotation elements, i.e. half wave plate (HWP) and quarter wave plate (QWP). A birefringent crystal (BC) couples the two DOF of a $p$-polarized photon in a fast pulse. The following combination of polarization elements, i.e. polarizer (POL), HWP and QWP set the outcome state.
}
\end{figure}

Polarization manipulation is achieved easily with standard wave plates and polarizers. In order to control the temporal DOF, first the photon's polarization is set, and than the polarization and the temporal DOF are coupled by passing through a birefringent crystal \cite{Shaham11}. The crystal length $L$ is set such that it induces a temporal polarization walk-off $\tau=\frac{L\Delta n}{c}$ much larger than the photon temporal pulse length, where $\Delta n$ is the birefringence index difference and $c$ the speed of light.

Consider a photon with the $|p\rangle$ polarization in time-bin $t=0$, traveling towards a birefringent crystal (see Fig. \ref{Fig1}). The two-qubit state is

\be
|\phi\rangle =|p,0\rangle=\frac{1}{\sqrt{2}}
\left(|h,0\rangle+|v,0\rangle\right).
\ee

Due to the crystal birefringence, the horizontal and vertical polarizations amplitudes, $|h\rangle$ and $|v\rangle$ travel at different group velocities through the crystal. Thus, upon entering the birefringent crystal, the photon $|h\rangle$ amplitude is not delayed, while the $|v\rangle$ amplitude is delayed by $\tau$. Thus, the photon exits the crystal in a superposition to be
non-delayed horizontally polarized and delayed vertically polarized. This results in the two-qubit maximally entangled Bell state

\be\label{phiplus} |\phi^+\rangle =\frac{1}{\sqrt{2}}
\left(|h,0\rangle+|v,\tau\rangle\right).
\ee
Starting with the orthogonal polarization $|\phi\rangle =|m,0\rangle$ would result in the orthogonal Bell
state
$|\phi^-\rangle=\frac{1}{\sqrt{2}}\left(|h,0\rangle-|v,\tau\rangle\right)$.

The process applied by the birefringent crystal for these two input states is actually an entangling gate between the two DOF. For three out of four input states this gate acts as a controlled-NOT (CNOT) gate. The exception is the $|v,\tau\rangle$ state which is delayed to $t=2\tau$ instead of accelerated to time-bin $t=0$, and thus transformed out the proper Hilbert space. Thus, this process in general is not unitary, nor trace preserving. Nevertheless, this process is sufficient for generating all of the states that we are interested in, as we will show below. The representation of this operator in the standard basis is:
\begin{equation}\label{gate}
\begin{aligned}
\hat{U}_{gate}&= \left( \begin{array}{cccc}
1 & 0 & 0 &  0  \\
0 & 1 & 0 &  0  \\
0 & 0 & 0 &  0  \\
0 & 0 & 1 &  0
\end{array}\right)
\end{aligned}
\end{equation}

\begin{figure}[t]
\includegraphics[width=\columnwidth]{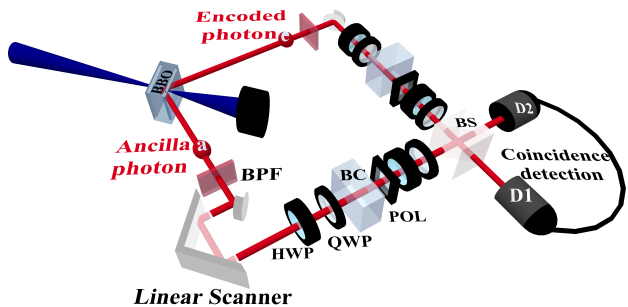}
\caption{\label{Fig2}The experimental setup.
Each photon is spectrally filtered using a bandpass filter (BPF).
An encoded photon is prepared in the desired state, while an ancilla photon is prepared in one of the desired projection states. The projection measurement is by bunching at the beam-splitter (BS), while single photon detectors (D1 and D2) record coincidence events as the arrival time of the ancilla photon is scanned by changing its path length using a linear motor scanner.}
\end{figure}

In order to generate an arbitrary polarization in a single time-bin, $t=0$ or $t=\tau$, the photon polarization state is set before the crystal to $|h,0\rangle$ or $|v,0\rangle$, accordingly (see Fig. \ref{Fig1}). A second set of polarization control elements is used to set the specific polarization. For a superposition of orthogonal states at the two time-bins, an initial $|p,0\rangle$ state before the crystal will create a balanced superposition after the crystal that can be rotated to any two orthogonal states. When a polarizer, which is also a non-unitary operation, is added after the crystal, two identical polarizations in the two time-bins are generated. The balance and the phase between the two amplitudes can be controlled by setting different polarization states before the crystal. If the polarization states of the two time-bins should be different, but non-orthogonal, a more elaborated setup is required, which is feasible, but outside the scope of the current work.

\section{III. CHARACTERIZATION METHOD}
As the temporal separation of photon amplitudes that is generated by birefringent crystals of a few millimeters is too small to detect directly, we present here another method that can characterize not just any time scale, but can also simultaneously detect any additional DOF. Assume a pure unknown state $|\psi_e\rangle$ encoded in a single photon and another ancilla photon at a known pure state $|\psi_a\rangle$. The two states differ by the vector $|\delta\rangle\perp|\psi_a\rangle$, such that $|\psi_e\rangle=\alpha|\psi_a\rangle+\beta|\delta\rangle$, where $|\alpha|^2+|\beta|^2=1$. It can be shown that if the two photons are injected into a balanced beam-splitter (BS) in a Hong-Ou-Mandel (HOM) \cite{HOM87} configuration where their relative arrival time is scanned, the ratio $R$ of the zero delay output coincidence rate to the long delay coincidence rate where the photons are completely distinguishable is
\be
R=|\beta|^2=1-|\alpha|^2=1-|\langle\psi_1|\psi_2\rangle|^2\,.
\ee
This projection is not restricted to encoded photons in unknown pure states, as the same ratio $R$ also applies for projections on mixed states where $|\delta_i\rangle\perp|\psi_a\rangle$
\be
\begin{aligned}
\rho_e=&|\alpha|^2|\psi_a\rangle \langle\psi_a|\\
+&\sum\limits_i \left( |\beta_i|^2|\delta_i\rangle \langle\delta_i| + \gamma_i|\delta_i\rangle\langle\psi_a|+H.C.\right)\,.
\end{aligned}
\ee
and $R=1-|\alpha|^2=1-\langle\psi_a|\rho_e|\psi_a\rangle$.
Thus, the projection of one photon state onto the other can be inferred from the HOM dip magnitude. Repeated measurements of a photon at an unknown state with an ancillary photon at well designed states can provide information for a full reconstruction of the unknown quantum state. This statement is mathematically valid regardless of the dimensionality of the states and the details of their DOF.

\section{IV. EXPERIMENTAL DEMONSTRATION}
The encoding and characterization of different polarization and temporal states was demonstrated using the experimental setup
presented in Fig. \ref{Fig2}. A pulsed Ti:sapphire laser source with a
$76\,$MHz repetition rate is frequency doubled by second harmonic generation
(SHG) to a wavelength of $390\,$nm and an average power of $300\,$mW. The
laser beam is corrected for astigmatism and focused on a $2\,$mm thick
$\beta$-BaB$_2$O$_4$ (BBO) crystal used for SPDC. The BBO angle is set at a beam-like configuration \cite{Kurtsiefer01,Takeuchi01}. The biphoton state is
spatially filtered by coupling the two beams into single-mode fibers, and spectrally
filtered using $3\,$nm bandpass filters at a wavelength of $780$\,nm. The resulting biphoton state is $|\psi\rangle=|h\rangle_e \otimes |v\rangle_a$, where $e$ represents the
\textit{encoded} photon and $a$ the \textit{ancilla} photon. The encoding setup used a $4\,$mm thick Calcite crystal that generates a time difference of $\tau=2.3\,$ps.


Using the above encoding procedure, we prepared the encoded photon in the following three two-qubit states
\be \label{3states}
\begin{aligned}
|\psi\rangle_{e,1}&=|\phi^+\rangle,\\
|\psi\rangle_{e,2}&=|p,+\rangle,\\
|\psi\rangle_{e,3}&=\frac{1}{\sqrt{2}} \left(|r,0\rangle-i|l,\tau\rangle\right).\\
\end{aligned}
\ee
These specific states were chosen as both DOF of $|\psi\rangle_{e,1}$ and $|\psi\rangle_{e,3}$ are entangled, while both the $|\psi\rangle_{e,2}$ and $|\psi\rangle_{e,3}$ states are mutually unbiased with the standard basis states, an important property for quantum key distribution protocols \cite{BB84}. The ancilla photon was prepared at a complete set of states, whose projections are sufficient for a full reconstruction of the encoded photon state using a quantum state tomography (QST) procedure \cite{James01}. A variable delay controlled the relative arrival time of the two photons to the HOM projecting BS. The delay introduces an additional distinguishability between the encoded and ancilla photons. The required projection values are obtained by recording the ratio $R$  of coincidence detection rates at the BS two output ports. If the ancilla state occupied a single time-bin, a single delay scan actually provided information for two projections.

\begin{figure}[t]
\centering
\subfigure[]{
\includegraphics[width=0.45\columnwidth]{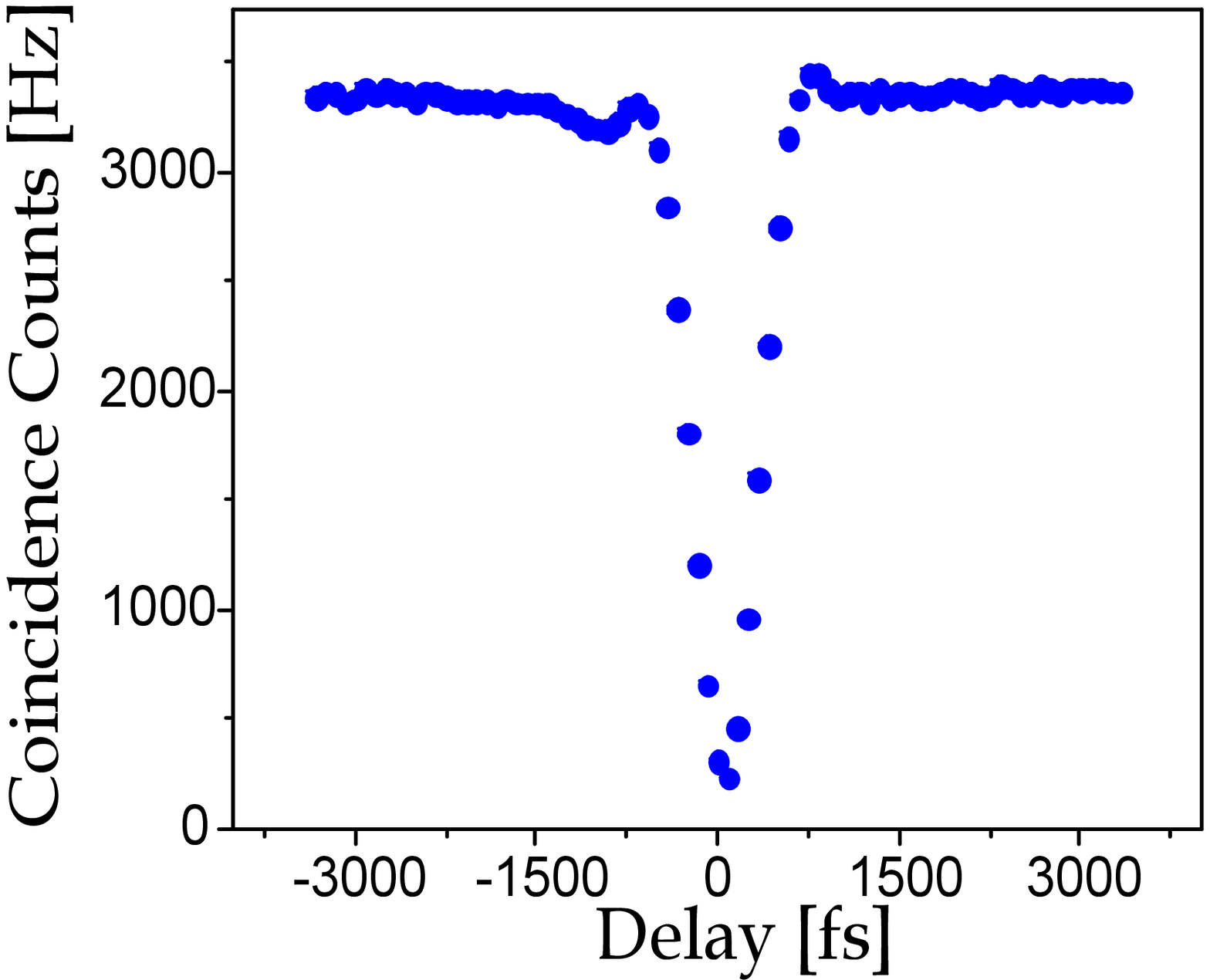}
\label{Fig3a}
}
\subfigure[]{
\includegraphics[width=0.45\columnwidth]{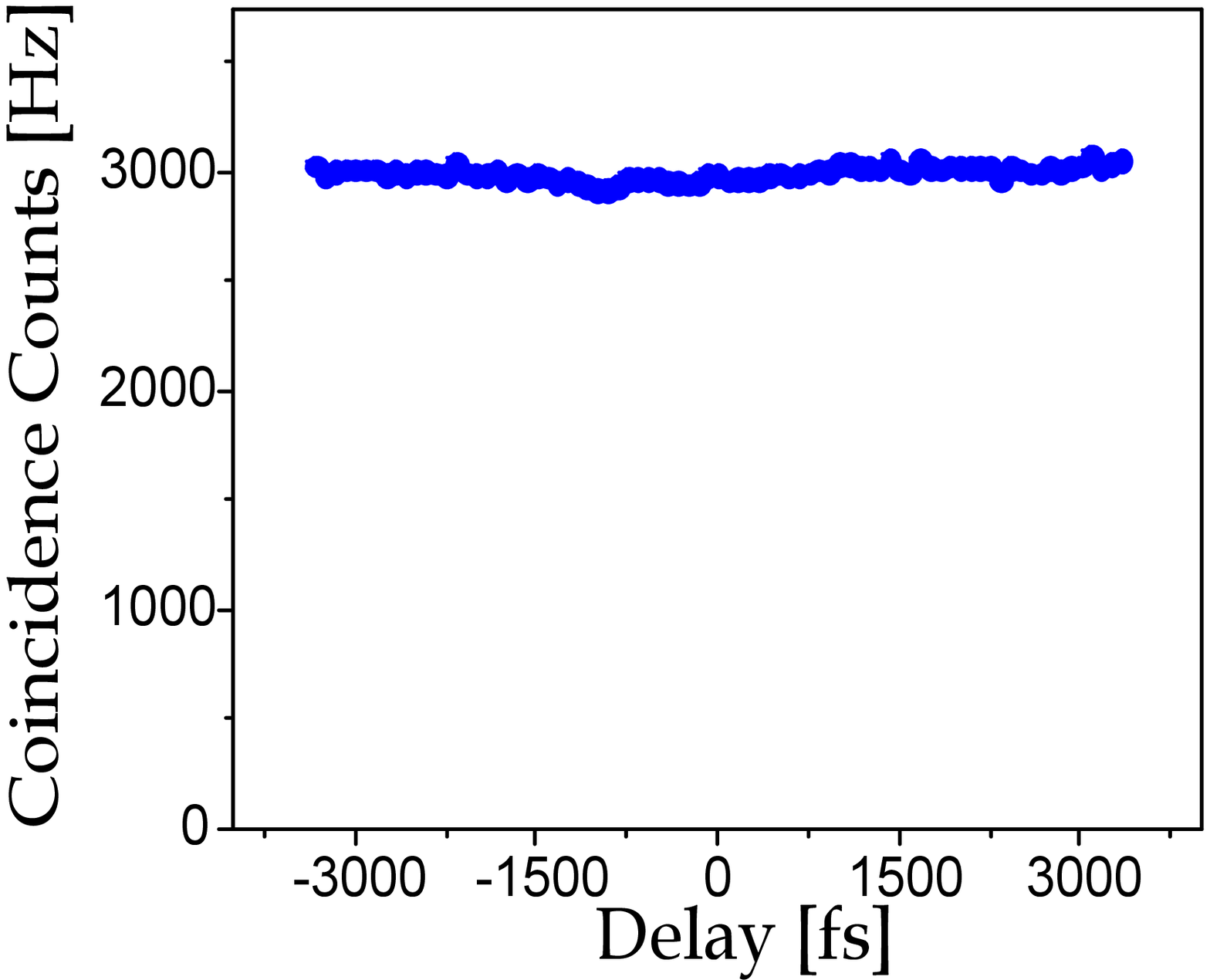}
\label{Fig3b}
}
\subfigure[]{
\includegraphics[width=0.45\columnwidth]{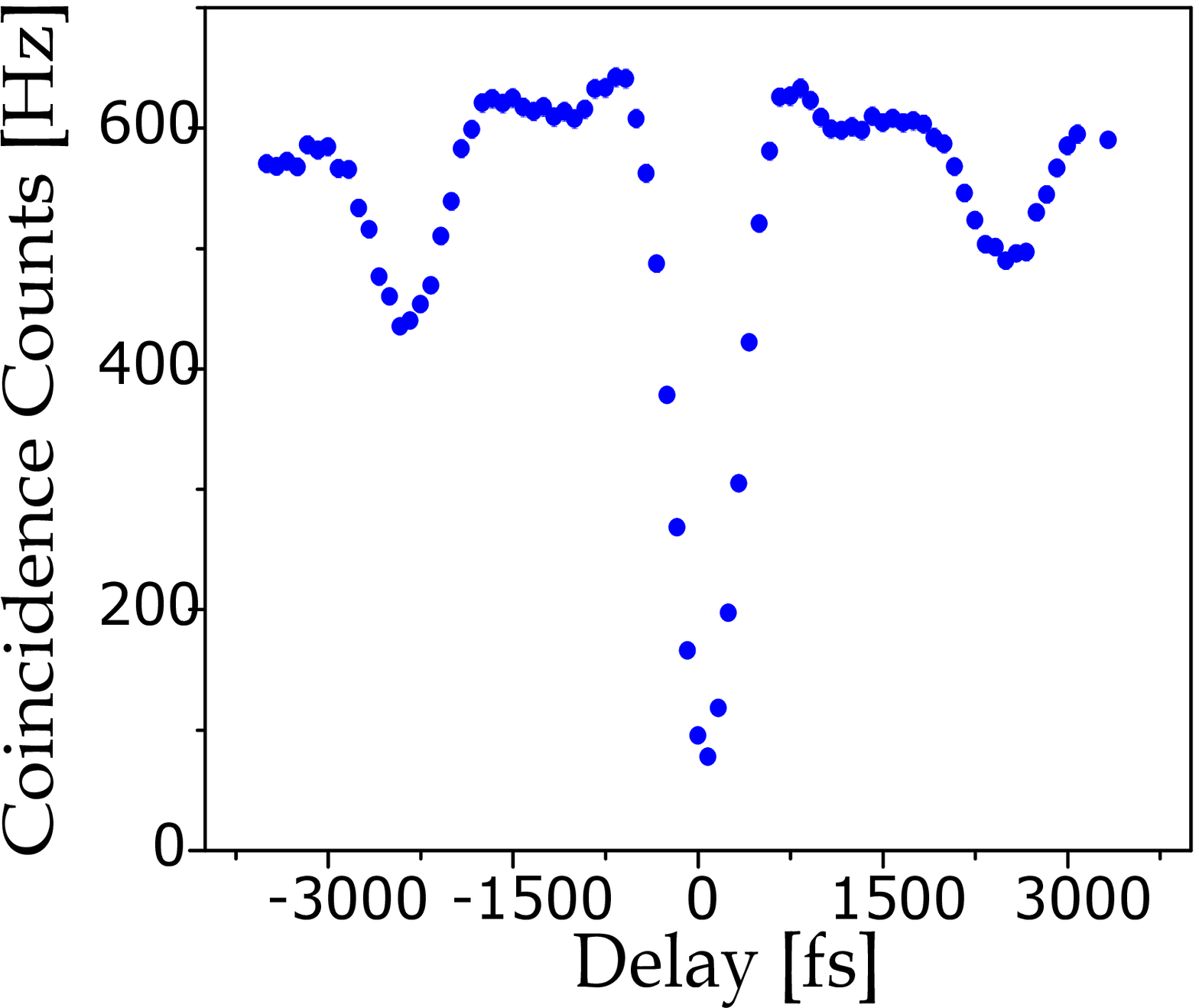}
\label{Fig3c}
}
\subfigure[]{
\includegraphics[width=0.45\columnwidth]{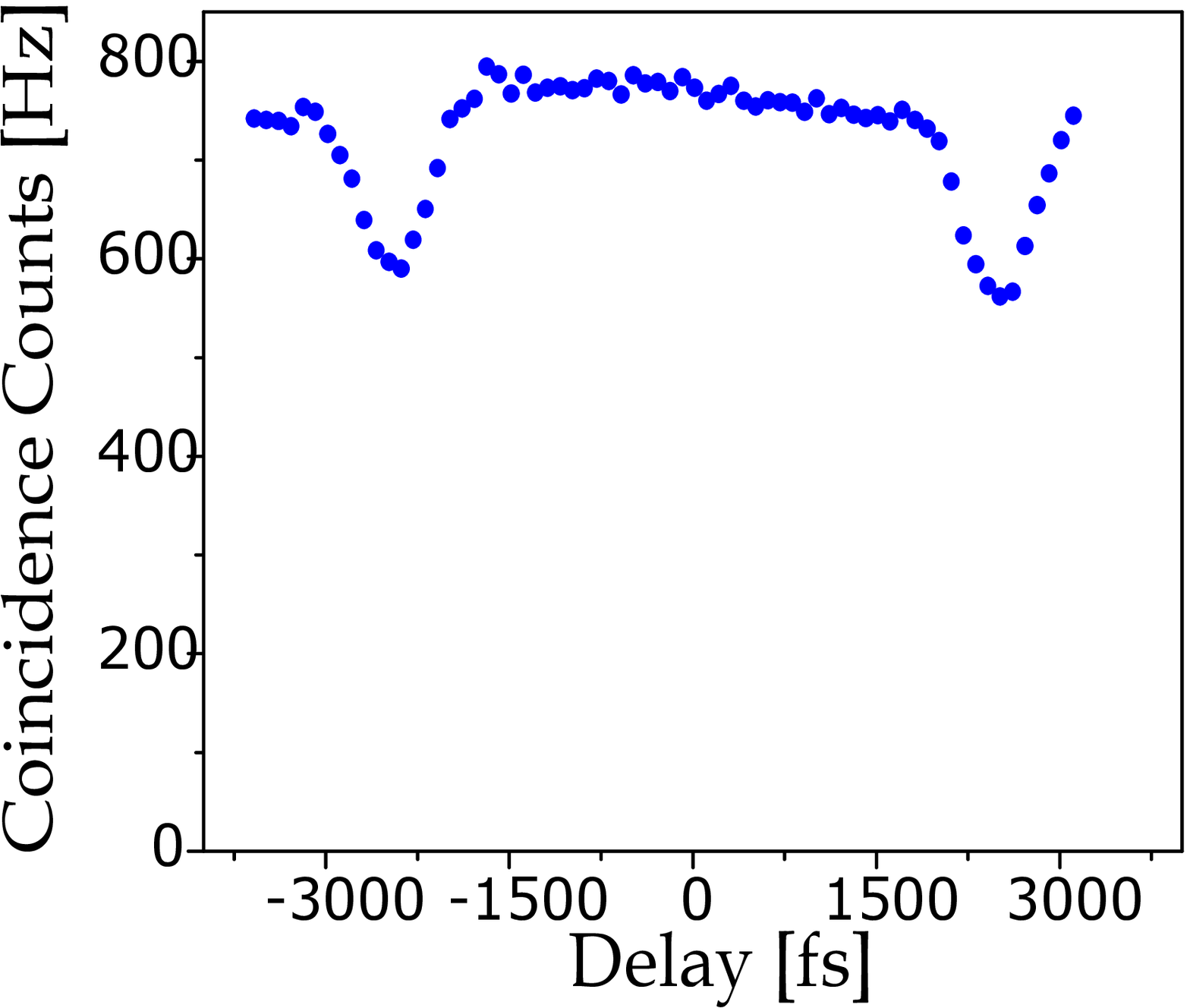}
\label{Fig3d}
}
\caption{\label{Fig3} Coincidence count rates as a function of the relative delay between the encoded and the ancilla photons. (a) Projecting $|\psi\rangle_{e,1}$ on ancilla photon at the same  $|\phi^+\rangle$ state and (b) on ancilla photon at the orthogonal $|\phi^-\rangle$ state. (c) Projecting $|\psi\rangle_{e,2}$ on ancilla photon at the same $|p,+\rangle$ state and (d) on ancilla photon at the orthogonal $|p,-\rangle$ state.}
\end{figure}

Figure \ref{Fig3} presents the experimental results of the coincidence detection rate of
the ancilla and the encoded photons at the output BS ports as a function of the introduced delay between them. When both photons are at the same state (Figs. \ref{Fig3a} and \ref{Fig3c}), a very good HOM interference dip visibility of $94\pm2\%$ and $89\pm2\%$ is observed at zero delay, respectively. When the photons are at orthogonal states (Figs. \ref{Fig3b} and \ref{Fig3d}), there is no dip at zero delay. The only difference between the scans of Figs. \ref{Fig3a} and \ref{Fig3b} is in the phase between the two time-bins, but it is sufficient to completely distinguish between the two states. The same phase difference is applied also between the scans of Figs. \ref{Fig3c} and \ref{Fig3d}, but here the polarization of the two time-bins is also identical. This results in a $1/4$ overlap of half of each state: at $-\tau$ delay the $t=\tau$ time-bin of the ancilla photon overlaps with the $t=0$ time-bin of the encoded photon and at $\tau$ delay the other two amplitudes overlap.

\begin{figure}[t]
\centering
\subfigure[]{
\includegraphics[width=0.45\columnwidth]{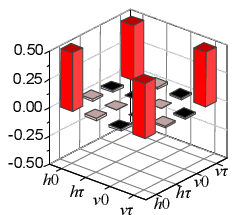}
\label{Fig4a}
}
\subfigure[]{
\includegraphics[width=0.45\columnwidth]{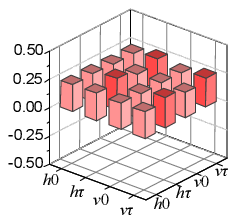}
\label{Fig4b}
}
\subfigure[]{
\includegraphics[width=0.45\columnwidth]{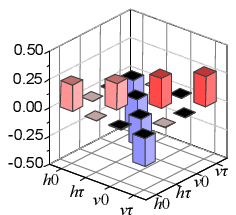}
\label{Fig4c}
}
\subfigure[]{
\includegraphics[width=0.45\columnwidth]{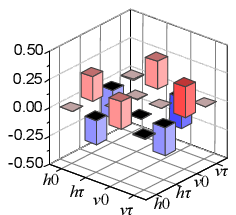}
\label{Fig4d}
}
\caption{\label{Fig4} Reconstructed density matrices of the encoded photons. (a) Real part of the measured $|\phi^+\rangle$ state. (b) Real part of the measured $|p,+\rangle$ state. The imaginary parts of (a) and (b) are not presented as their values are zeroes within the experimental accuracy. (c,d) The real and imaginary parts of the measured $|\psi\rangle_{e,3}$ state, respectively.}
\end{figure}

The three states of Eq. \ref{3states} were fully characterized by the QST procedure \cite{James01}. For each encoded state, projection on a complete set of 16 states provided the required information for the QST procedure, where physically valid results were obtained utilizing a maximum likelihood procedure. The reconstructed density matrices are presented in Fig. \ref{Fig4}. High fidelities with the theoretical states of Eq. \ref{3states} were observed \cite{Josza94}. The generated $|\psi\rangle_{e,1}$, $|\psi\rangle_{e,2}$, and $|\psi\rangle_{e,3}$ states had $98\%\pm2\%$, $96\%\pm1\%$, and $98\%\pm2\%$ fidelities with their respective theoretical states, showing a very good agreement between theory and experiment. Errors were calculated by propagating Poissonian statistical errors in the coincidences rates with a bootstrap approach through the maximum likelihood QST protocol.

\section{V. CONCLUSIONS}
In conclusion, we present a simple scheme to utilize simultaneously the polarization
and time DOF of a single photon, realizing a two-qubit information system. Instead of using imbalanced interferometers which are large and unstable, we used birefringent crystals to generate the temporal information. As the temporal details generated in this method are too fine for direct detection, we suggested and demonstrated a new characterization method, based on the HOM effect. This method characterizes all of the DOF simultaneously, without the need for separate optical setups for each DOF. The dimensionality of the generated states can be extended in a straight forward manner by adding more birefringent crystals. The presented high quality results support the feasibility of the suggested methods for encoding fine temporal information, and decoding quantum information of several DOF efficiently.

\section{ACKNOWLEDGMENTS}
The authors thank the Israeli Science Foundation for supporting
this work under grants 793/13 and 2085/18.

Y.P. and P.Z. contributed equally to this work.

\end{document}